%% file: 2020-TimeDetWirelessStack-Preprint.tex
\documentclass[journal]{IEEEtran}

\ifCLASSOPTIONcompsoc
  \usepackage[nocompress]{cite}
\else
  \usepackage{cite}
\fi

\usepackage{array}
\usepackage{stfloats}
\usepackage{graphicx} 
\usepackage{times} 
\usepackage{amsmath}
\usepackage{amssymb}
\usepackage{mathptmx}
\usepackage{color}
\usepackage{url}
\usepackage{algorithm}
\usepackage{algorithmic}
\usepackage{multirow}
\usepackage{caption}
\usepackage{booktabs}
\usepackage[para,online,flushleft]{threeparttable}
\usepackage{tabularx}
\usepackage[acronym]{glossaries}
\usepackage{listings}

\newacronym{WSN}{WSN}{Wireless Sensor Network}

\usepackage[colorinlistoftodos, textwidth=14mm, textsize=footnotesize]{todonotes}
\usepackage{enumitem}
\setlist{nolistsep}

\usepackage{booktabs} 

\usepackage{tikz,xstring,siunitx,pgfplots}
\usetikzlibrary{shapes,arrows}
\usetikzlibrary{calc,patterns,decorations.pathmorphing,decorations.markings}
\usetikzlibrary{circuits,circuits.ee.IEC}
\usetikzlibrary{pgfplots.groupplots}

\tikzstyle{block} = [draw, rectangle, minimum height=2em, minimum width=2em]
\tikzstyle{sum} = [draw, circle, node distance=1.5cm]
\tikzstyle{input} = [coordinate]
\tikzstyle{output} = [coordinate]

\pgfplotsset{compat=newest}

\usepgfplotslibrary{external} 
\makeatletter
\renewcommand{\todo}[2][]{\tikzexternaldisable\@todo[#1]{#2}\tikzexternalenable}
\makeatother

\usepackage[nolist,nohyperlinks]{acronym}
\acrodef{WSN}[WSN]{Wirless Sensor Network}

\usepackage{color}
\definecolor{mygreen}{rgb}{0,0.6,0}
\definecolor{mygray}{rgb}{0.5,0.5,0.5}
\definecolor{mymauve}{rgb}{0.58,0,0.82}

\begin{document}

\title{{TDMH: a communication stack\\for real-time wireless mesh networks}}

\author{Federico~Terraneo,
        Federico~Amedeo~Izzo,
        Alberto~Leva,
        and William~Fornaciari

        \IEEEcompsocitemizethanks{
        \IEEEcompsocthanksitem
          F. Terraneo, A. Leva and W. Fornaciari are with the DEIB, Politecnico di Milano, Italy.
          E-mail: \{federico.terraneo,alberto.leva,william.fornaciari\}@polimi.it
        \IEEEcompsocthanksitem
          F.A. Izzo is a former graduate students at the Politecnico di Milano.
          E-mail: \{federico.izzo\}@mail.polimi.it
          }
}

\markboth{}%
{Terraneo \MakeLowercase{\textit{et al.}: {TDMH-MAC: a real-time multi-hop WSN MAC protocol}}}

\IEEEtitleabstractindextext{%
\begin{abstract}
\input{./sections/00-Abstract.tex}

\end{abstract}

}

\maketitle
\IEEEdisplaynontitleabstractindextext
\IEEEpeerreviewmaketitle
\acresetall

\input{./sections/01-Introduction.tex}
\input{./sections/02-RelatedWork.tex}

\input{./sections/03-ProposedProtocol.tex}

\input{./sections/04-TopologyCollection.tex}

\input{./sections/05-Scheduler.tex}

\input{./sections/07-SimScalability.tex}

\input{./sections/08-ExpRes.tex}

\input{./sections/09-Conclusions.tex}

\bibliographystyle{IEEEtran}

\end{document}

%% file: sections/00-Abstract.tex
We present the TDMH (Time Deterministc Multi-Hop) protocol, a complete stack for real-time wireless mesh networks. TDMH offers to applications a connection-oriented, bounded-latency communication model. Point-to-point data streams can be created and destroyed at any time. Path redundancy can be optionally introduced to improve reliability. TDMH exploits state-of-the-art low power clock synchronisation and constructive interference flooding to build a continuously updated graph of the network topology, onto which a centralized scheduler maps data streams using TDMA channel access. We realised TDMH as a unitary codebase, that we ran on both the OMNeT++ simulator and WandStem wireless nodes. As a result we can state that when built atop the IEEE 802.15.4 physical layer, TDMH can scale up to 100 nodes, 10 hops and beyond, despite the limited available bandwidth.

%% file: sections/01-Introduction.tex
\section{Introduction}
\label{sec:Intro}

The landscape of wireless protocols is constantly evolving to address the growing need for connectivity in the modern world.
Existing wireless solutions fulfill the requirements of many application areas, and are widely deployed in the industry.
However, recent directions such as Industry 4.0, the industrial IoT and Cyber Phisical Systems (CPS)~\cite{7945906,7993826}, call for real-time wireless networks, which is an area still in search of major breakthroughs to reach wide adoption.

The main reason that singles out real-time networks is that time determinism - unlike other network properties - cannot be improved through layering.
Let us discuss this point in more detail. Properties such as reliability can be improved by adding a layer that adds checksums and performs retransmissions if said checksum don't match. Limitations in the maximum transmissible packet size can be overcome through a fragmentation layer. Many more examples could be given where layering allows to overcome a network limitation.
However, since time cannot be made to run backwards, a time deterministic protocol cannot be built by adding another layer to a less time deterministic protocol.

This remark is our main motivation for the development of Time Deterministic Multi-Hop (TDMH), a new protocol stack that has been designed from the ground up for real-time networks.
TDMH targets the market of wireless sensors and actuators for industrial automation and CPS in general.
Due to its low energy demand, battery operated devices are another natural target.

TDMH differs in several key aspects from mainstream network stacks. To achive real-time communication, data frames are transmitted without packet queues and are not acknowledged. Reliability is instead ensured through redundant transmissions with preallocated bandwidth. TDMH uses constructive interference flooding both for network management and to achieve low overhead clock synchronization using the FLOPSYNC-2 protocol~\cite{bib:TerraneoEtAl-2014a}.
Synchronization is natively exposed to the operating system, making it possible for applications and communication to operate in concert, enabling time deterministic distributed embedded systems.

The TDMH stack provides functionalities spanning from the Data Link to the Session layers of the ISO/OSI model.
The MAC layer of TDMH constructs a graph of the mesh network using an innovative distributed algorithm, and keeps it constantly updated.
Data transmissions can follow redundant paths exploiting the mesh topology for reliability.
A centralized routing and scheduling algorithm coordinates the network, enabling collision-free deterministic communication with bounded latency.
If the network topology changes, the schedule is recomputed, allowing for adaptation and robustness to link and node failure.
The centralized network management also allows to reuse individual time slots for independent non-interfering transmissions, greatly increasing the bandwidth for data communication.
TDMH exposes to applications a communication model based on \emph{streams}. A stream is a unidirectional or bidirectional data flow that can be opened at runtime between any two nodes in the network. The individual data unit in a stream is the packet, which are transmitted with a guaranteed period and with bounded latency.

To the best of the authors' knowledge, TDMH is the first network stack providing a holistic solution to the real-time wireless mesh networking problem, where each component has been designed specifically for time determinism and the overall goal is achieved through the interplay of its parts.

TDMH has been validated both on the OMNeT++ network simulator and with real sensor nodes. The reference implmentation consists in a unitary codebase released as free software\footnote{https://github.com/fedetft/tdmh}. TDMH can scale to networks of more than 100 nodes and 10 hops.

%% file: sections/02-RelatedWork.tex
\section{Related Work}
\label{sec:RelatedWork}

Dedicated wireless protocols for mesh networks and/or real-time applications have been proposed by several research communities as well as from the industry.
One such research community is the Wireless Sensor Network one.
There, the need for low power protocols operating with limited computational capabilities has favored custom protocols instead of a standardization around general-purpose solutions based on TCP/IP.
A taxonomy of WSN protocols~\cite{wsnmacsurvey} shows they are divided in four branches: asynchronous, synchronous, frame-slotted, and multichannel.
Asynchronous and synchronous protocols suffer from poor time determinism caused by the use of CSMA/CA, while frame-slotted protocols are based on TDMA. Among frame-slotted protocols, TreeMAC~\cite{treemac}, PackMAC~\cite{packmac} and TRAMA~\cite{trama} are examples targeted to convergecast traffic patterns, employing spanning tree routing.
Multichannel protocols reduce contention and improve throughput by operating on multiple channels. Examples include MMSN~\cite{mmsn} which is based on a common control channel, and separate ones for data transmission, while MuChMAC~\cite{muchmac} relies on channel hopping.
Although the mentioned protocols support multi-hop networks, they do not provide delay bounds unlike TDMH. Moreover, some of them have restrictions in the supported traffic patterns, such as being limited to convergecast.
WSN protocols supporting real-time communication are uncommon, but have been proposed. (RT)$^{2}$~\cite{4453854} builds a real-time protocol for WSN networks on top of a standard CSMA/CA MAC, by adapting the data rate based on congestion. While this approach can maximize throughput, application transmission periods are not guaranteed.


Mesh network protocols is an active research topic also outside WSNs, with works dedicated entirely to providing design considerations for wireless mesh networks~\cite{5418856}.
The work by Cheng et al.~\cite{4560038} introduces a routing protocol for large scale mesh network taking advantage of directional communication. The solution avoids flooding, which is beneficial for very large scale mesh networks, but has the disadvantage to produce suboptimal routes. TDMH, being targeted to smaller industrial-scale networks takes advantage of constructive inteference-based flooding, which scales easily to over a hundred nodes.
The work by Kim et al.~\cite{5678602} proposes a framework that automatically reconfigures IEEE 802.11 radios to overcome common causes of link failures in mesh networks. This approach would be benefical also in TDMH networks, but is currently outside the scope of this paper.


Several wireless protocols are part of the IEEE 802.15.4 standard~\cite{802154e-survey}, three of which are relevant for their mesh or real-time capabilities.
TSCH is a protocol combining TDMA and multichannel support. It is based on a cluster-tree topology where time is divided in repeating slotframes with a common schedule.
DSME~\cite{OpenDSME} exploits the IEEE 802.15.4 Collision Free Part (CFP) to guarantee bandwidth allocation.
Slot allocation is managed in a distributed way, allowing each node to autonomously allocate or deallocate slots, resulting in a different architecture with respect to TSCH.
LLDN is specifically designed for low latency real-time applications, however, it only supports a star topology~\cite{7005204}.
TDMH differs from these protocols by providing delay bounds on a multi-hop network, by taking advantage of a mesh topology, allowing for redundant paths to enhance reliability, and by improving channel utilization thanks to constructive interference flooding.


Real-time wireless networks can also be built atop consumer oriented physical layers, such as Bluetooth Low Energy~\cite{8355905} or Wi-Fi.
RT-WiFi~\cite{6728869} transmits TDMA-scheduled packets enjoying the high data rate of Wi-Fi, but is limited to a star topology. Other approaches try to cope with the limitations of CSMA, by using Quality of Service (QoS)~\cite{7389381} to prioritize real-time traffic, or jamming the channel first to stop non real-time traffic and then transmitting, a technique referred to as bandjacking~\cite{8336984}.
The work by Yang et al.~\cite{6841643} proposes a custom PHY layer that adapts the network coding at runtime to meet soft deadlines, while active drop queues~\cite{1709950} have been introduced to support bursty soft real-time application, such as multimedia traffic.


Industrial wireless protocols for factory automations also exist. These are typically propietary protocols, whose specifications are not openly available.
WirelessHART\cite{wirelesshart} and ISA-100.11a~\cite{ISA-100.11a} are the two industry standards. As illustrated in \cite{whartvsisa} they both provide a TDMA protocol and use a mesh topology, but unlike TDMH do not provide latency bounds~\cite{ISA100.11a-closedloop,WirelessHart-closedloop}, a matter that complicates closing control loops encompassing wireless links. Moreover, said protocols do not allow to reuse time slots for concurrent non-interfering transmissions, thereby using the channel bandwidth less efficiently than TDMH.


The closest approach to TDMH in terms of features is the work by Mager et al~\cite{bib:MagerEtAl-2019a} which proposes a wireless embedded system which transmits every data packet using constructive interference flooding, thus transforming a multi-hop network in a single broadcast domain. This approach is however not yet a general purpose protocol, as the transmission schedule is hardcoded, and any transmission change would require to reprogram the firmware of all the wireless nodes in the network.

Table~\ref{tab:mac-comparison} compares the features of TDMH to the most relevant protocols whose specifications are openly available.

\begin{threeparttable}[b]
\vspace{4mm}
\caption{Comparison of existing wireless MAC protocols.}
{\footnotesize \centering
		\begin{tabularx}{\columnwidth}{l|ccccc}
			Feature             & TDMH      & TSCH         & DSME       & LLDN       & rt-WiFi\\
			\hline
			Multi-hop           & \checkmark & \checkmark   & \checkmark &            &  \rule{0pt}{2.6ex}\\ [1mm]
				Bounded          & \multirow{2}{*}{\checkmark} & \multirow{2}{*}{} & \multirow{2}{*}{} & \multirow{2}{*}{\checkmark} & \multirow{2}{*}{\checkmark} \\
                latency \\ [1mm]
				Spatial             & \multirow{2}{*}{\checkmark} & \multirow{2}{*}{} & \multirow{2}{*}{} & \multirow{2}{*}{} & \\
				redundancy \\ [1mm]
				Temporal            & \multirow{2}{*}{\checkmark} & \multirow{2}{*}{\checkmark} & \multirow{2}{*}{feasible} & \multirow{2}{*}{feasible} & \multirow{2}{*}{\checkmark} \\
				redundancy \\ [1mm]
			Management          & C\tnote{1} & C\tnote{1}/D\tnote{2} & D\tnote{2} & C\tnote{1} & C\tnote{1} \\ [1mm]
			Topology            & mesh       & ct\tnote{3} & ct\tnote{3} & star & star \\ [1mm]
		\end{tabularx}
		\vspace{0.4em}
		\begin{tablenotes}
            \item[1] centralized \item[2] distributed \item[3] cluster-tree
        \end{tablenotes}
        \vspace{0.4em}
	\label{tab:mac-comparison}
}
\end{threeparttable}

As can be seen from Table~\ref{tab:mac-comparison}, existing protocols can be divided between those guaranteeing tight latency bounds, which are however limited to star topologies, and protocols supporting multi-hop networks, where providing any form of latency bound is a much more uncommon feature.
For what concerns the latter, network topology is usually limited to a cluster-tree to overcome the difficulties in discovering the network topology.
Such a solution removes a-priori some links which could be used to improve reliability. TDMH innovates in this respect by providing an efficient solution to the topology collection problem.
Moreover, when considering centralized protocols, TDMH is, to the best of the authors' knowledge, the first using constructive interference flooding to disseminate routing information and clock synchronization.

%% file: sections/03-ProposedProtocol.tex
\section{TDMH Protocol Design}
\label{sec:Protocol}

\begin{figure}[tb]
 \begin{center}
  \includegraphics[width=\columnwidth]{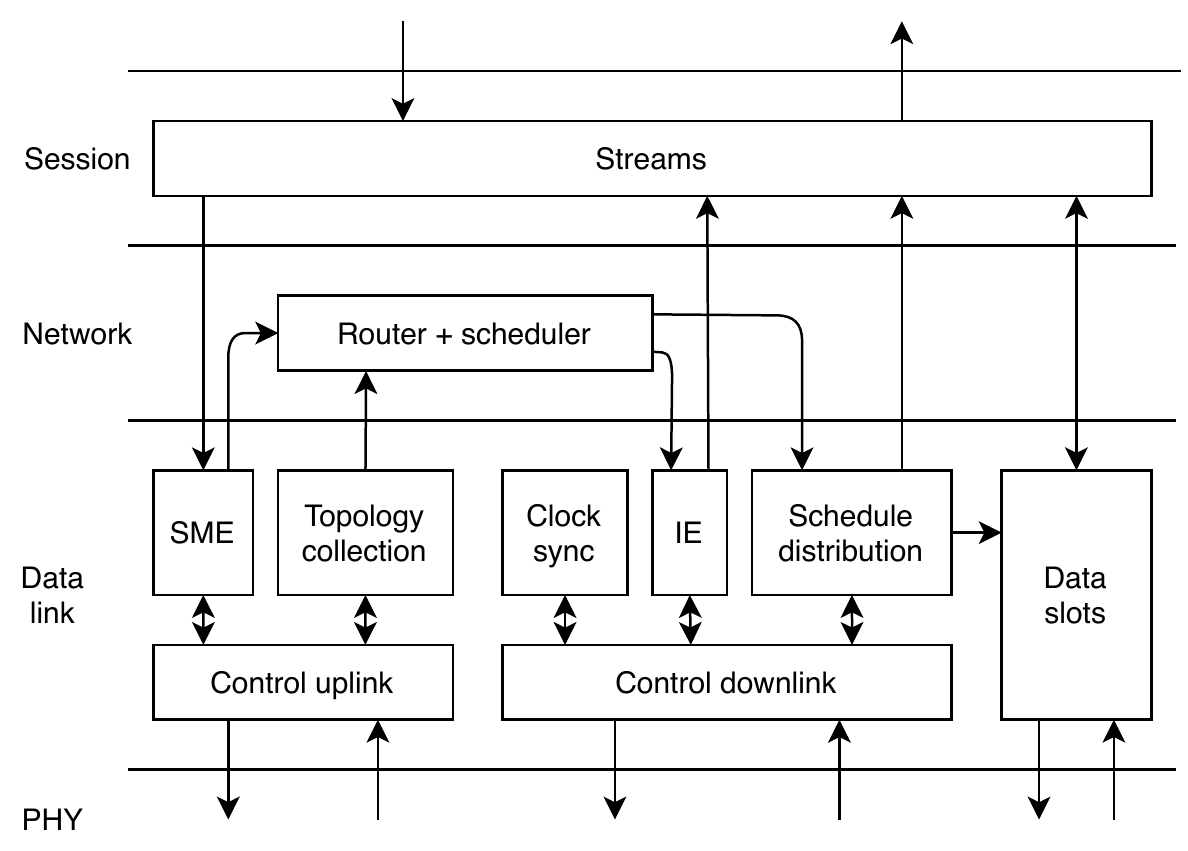}
 \end{center}
 \caption{Overview of the TDMH networking stack.}
 \label{fig:TDMH-components}
\end{figure}

TDMH is a centralized, connection-oriented networking stack. It targets periodic real-time communication applications.
We apologize in advance to the reader for the density of the following sections, but the holistic nature of the TDMH, coupled with its design differences compared to other networking stacks require it to be described with significant level of detail to understand its operation.

An overview of TDMH is shown in Figure~\ref{fig:TDMH-components}.
The data link layer is run by all nodes in the network, and uses TDMA to transmit \emph{data} frames and \emph{control} frames, which are used for network synchronization and management.
The network layer is composed of a global router and scheduler. This layer is present only in one node which operates as master, usually the gateway.  All nodes except the master have to communicate with the scheduler through the network, hence the direct connection between the session and data link layers.
The session layer handles the streams logic, with stream opening and closing requests being forwarded to the scheduler, and application data being directly encapsulated in data frames.

\subsection{PHY layer interface}
\label{sec:phylayer}

The TDMH data link layer accesses the physical layer through only four Service Access Point (SAP) primitives, shown in Listing~\ref{physap}, plus additional PHY-specific ones to perform operations such as transceiver configuration. Two peculiar features of TDMH are worthy of note.
\begin{lstlisting}[caption={PHY SAP primitives required by TDMH.},captionpos=b,label=physap]
PHY-TX.request(frame,sendTime)
PHY-TX.confirm(errorCode)
PHY-RX.request(timeout)
PHY-RX.confirm(errorCode,frame,timestamp)
\end{lstlisting}
First, the packet transmission primitive specifies not only the frame to be sent, but also the absolute time (in the future) when it must be sent. The resolution and jitter of the packet transmission, as well as the packet reception timestamping must be accurate enough to allow constructive interference. For the IEEE 802.15.4 protocol, the total end-to-end jitter must be less than 500ns~\cite{bib:FerrariEtAl-2011a}, and in our implementation the transmission jitter is 21ns~\cite{bib:Terraneo:2016:DHE:2893711.2893753}, while the receive jitter is in the order of 50ns~\cite{bib:TerraneoEtAl-2014a}, thus allowing to reliably achieve constructive interference.

Second, the TDMA nature of the MAC is specifically designed for real-time operation as well as natively supporting low power operation.
No packet queues are required between the PHY and MAC, with the PHY-TX and PHY-RX SAP only handling a single outstanding packet. All frames are transmitted without a corresponding acknowledge frame, and there is no retransmission logic. 

For what concerns power consumption, it must be noted that the MAC layer performs an explicit receive request in all TDMA slots where a packet is expected to be received. Thus the PHY layer does not have to continuously listen for possible packets and can natively go in a deep sleep state as much as possible, resulting in the power consumption of each node scaling linearly with the bandwidth of communications that traverse it.

\subsection{Data-link layer}

The TDMH data link layer can be logically viewed as composed of three distinct activities, control downlink, control uplink and data transmission.
The network control bandwidth is statically assigned network wide by dedicating certain transmission slots for control frames, in order to leave a guaranteed bandwidth for real-time data communication. The possibility to statically tune the control bandwidth as part of the network configuration allows to trade off faster reaction to topology changes and faster stream opening/closing with greater bandwidth for data frames.

\subsubsection{Temporal organization}

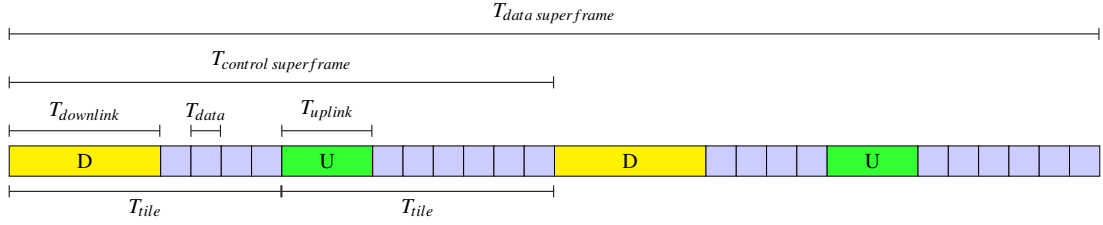
\begin{figure*}[tb]
 \begin{center}
  \resizebox{1.65\columnwidth}{!}{\input{./img/MACstructure.tex}}
 \end{center}
 \caption{Temporal organization of the TDMH MAC (not to scale). Transmission is organized in tiles of two types but of equal duration. Downlink tiles start with a flooded frame from the master, Uplink tiles start with a broadcast transmission by one of the nodes.}
 \label{fig:ProtocolStructure}
\end{figure*}

The MAC layer of TDMH is temporally organized as shown in Figure~\ref{fig:ProtocolStructure}. The protocol is organized in \emph{tiles}. All tiles begin with a control slot, where one control transmission, either downlink or uplink, occurs. The rest of the tile is occupied by data slots where data frames are transmitted. All tiles are the same duration, and since downlink slots are larger due to the need to flood a frame across multiple hops, downlink tiles have fewer data slots.

The shortest repeating sequence of downlink and uplink tiles is called a control superframe. A control superframe must have at least one downlink and one uplink tile for the MAC to be able to perform both activities.

The schedule computed by the master node defines the data superframe, whose duration is necessarily a multiple of the control superframe due to the asymmetry in the number of data slots of the different tiles. While the control superframe is a fixed configuration parameter, the data superframe duration changes at runtime depending on the current schedule.

The regular structure of TDMH tiles makes it possible for data streams to not only have guaranteed bandwidth, but also guaranteed period and bounded latency, which are important for periodic real-time communication.

\subsubsection{Control downlink}

Control downlink frames always originate from the master node and are flooded to the entire network using a Glossy-like~\cite{bib:FerrariEtAl-2011a} constructive interference-based flooding scheme. Control downlink frames are used for three purposes: for network synchronization using the FLOPSYNC-2~\cite{bib:TerraneoEtAl-2014a} clock synchronization scheme, to disseminate schedules, and to send Information Elements (IE) which are used to manage stream opening/closing requests.

\subsubsection{Control uplink}

Control uplink frames allow nodes to communicate with the master. Uplink frames are broadcast but not flooded, and are thus received only by each nodes' direct neighbors, a feature exploited to construct the network graph. Control uplink frames contain network topology information as well as Stream Management Elements (SME) which are used for stream opening/closing requests.
Topology and SME information are stored in queues and relayed in subsequent uplink slots following a convergecast pattern, finally reaching the master node where they are processed by the router and scheduler.
Access contention for control uplink slots is solved using a round-robin scheme.

\subsubsection{Data slots}

Data slots are used to transmit data frames from streams according to the global TDMA schedule. The operation of the MAC in the data slots is entirely schedule-driven. Each node knows in advance the action to perform for each data slot in the schedule, which can be one of the following:
\begin{itemize}
\item transmit a frame containing data from the session layer,
\item receive a frame and store it in a local buffer,
\item forward a buffered frame,
\item receive a frame and pass it to the session layer,
\item or sleep, saving power.
\end{itemize}
The scheduler can schedule multiple independent transmissions (having different source and destination nodes) in the same TDMA slot, as long as they do not interfere, a feature that in TDMH is called \emph{channel spatial reuse}. This bandwidth-enhancing feature is made possible by the full mesh topology knowledge as well as by the centralized network management.

\subsection{Network layer}

At the TDMH network layer we find the global router and scheduler operated by the master node. The router uses the information provided by the topology collection distributed algorithm to find paths in the mesh network that connect the endpoint nodes of each stream. The scheduler globally schedules data transmissions in specific data slots resulting in no access contention and thus guaranteeing the periods and latency of streams.
The scheduler also acts as an admission control preventing real-time operation issues due to network overload.
A more detailed explanation of the routing and scheduling algorithm is given in Section~\ref{sec:sched}.

\subsection{Session layer}

At the session layer we find streams, which are the abstraction that TDMH exposes to applications. Streams are by design similar to TCP sockets to leverage application developer familiarity, with the main difference that streams have the knowledge of a transmission period and only allow up to one packet to be transmitted per period, fulfilling the periodic real-time design goals.

There are two types of streams: server and client. Server stream do not participate in data transmission, but can only \texttt{listen} on a specific port and \texttt{accept} connections. A successful accept operation returns a client stream that can be used for communication. To initiate communication, a node has to perform a \texttt{connect} operation to a (node, port) tuple with a matching server.

Given the centralized nature of the protocol, stream opening/closing requests are not exchanged directly between endpoints, but forwarded to the router and scheduler as Stream Management Element (SME) data structures sent in control uplink frames. Upon acceptance of a new data stream, a new schedule is transmitted, implicitly marking the beginning of data exchange between endpoints. Server streams as well as failed client stream open requests are notified using Information Elements (IE) transmitted in control downlink frames.
A stream opeining request includes a number of parameters, which are
\begin{itemize}
\item the transmission period,
\item whether the stream is bidirectional or unidirectional, and in the latter case the direction
\item the redundancy requirement.
\end{itemize}

As already stated, all frames in TDMH, including data frames are sent without acknowledge and retransmissions, which implies the possibility of packet losses due to external interference. To mitigate this issue, a redundancy level can be requested on a per-stream basis, allowing each application packet to be unconditionally scheduled for independent transmission up to three times within each period, trading off bandwidth for reliability.
In addition, exploiting the mesh network information, a peculiar feature called \emph{spatial redundancy} can be requested, in which case the redundant data frames are routed following two independent paths in the network, when the topology makes it possible. Spatial redundancy can mitigate link or intermediate node failures, allowing streams to continue operation while the topology collection algorithm detects the change and a new schedule is computed and disseminated.

It should however be noted that TDMH, to meet its real-time design goals, does not guarantee the delivery of each application packet. According to the experimental results presented later on, the achieved reliability is very high, above 99\%. Such a situation can be accepted even in highly critical applications such as real-time feedback control~\cite{seiler2001analysis}.

%% file: img/MACstructure.tex
\begin{tikzpicture}

\node[right,draw=black,fill=yellow,
      minimum width=25mm,minimum height=5mm]
      at (0,0) (D1) {D}; 
\foreach \x in {25mm,30mm,35mm,40mm}
{
\node[right,draw=black,fill=blue!20!white,
      minimum width=5mm,minimum height=5mm]
      at (\x,0){};
}
\node[right,draw=black,fill=green!80!white,
      minimum width=15mm,minimum height=5mm]
      at (45mm,0) (U1) {U}; 
\foreach \x in {15mm,20mm,25mm,30mm,35mm,40mm}
{
\node[right,draw=black,fill=blue!20!white,
      minimum width=5mm,minimum height=5mm]
      at (\x+45mm,0){};
}

\node[right,draw=black,fill=yellow,
      minimum width=25mm,minimum height=5mm]
      at (90mm,0) (D2) {D}; 
\foreach \x in {25mm,30mm,35mm,40mm}
{
\node[right,draw=black,fill=blue!20!white,
      minimum width=5mm,minimum height=5mm]
      at (\x+90mm,0){};
}
\node[right,draw=black,fill=green!80!white,
      minimum width=15mm,minimum height=5mm]
      at (135mm,0) (U2) {U}; 
\foreach \x in {15mm,20mm,25mm,30mm,35mm,40mm}
{
\node[right,draw=black,fill=blue!20!white,
      minimum width=5mm,minimum height=5mm]
      at (\x+135mm,0){};
}

\draw[|-|] (0,-5mm)
   -- node[pos=0.5,below]{$T_{tile}$}
  (45mm,-5mm);
\draw[|-|] (45mm,-5mm)
   -- node[pos=0.5,below]{$T_{tile}$}
  (90mm,-5mm);
\draw[|-|] (0,5mm)
   -- node[pos=0.5,above,yshift=0.5mm]{$T_{downlink}$}
  (25mm,5mm);
\draw[|-|] (30mm,5mm)
   -- node[pos=0.5,above,yshift=0.5mm]{$T_{data}$}
  (35mm,5mm);
\draw[|-|] (45mm,5mm)
   -- node[pos=0.5,above,yshift=0.5mm]{$T_{uplink}$}
  (60mm,5mm);
\draw[|-|] (0,13mm)
   -- node[pos=0.5,above,yshift=0.5mm]{$T_{control\;superframe}$}
  (90mm,13mm);
\draw[|-|] (0,21mm)
   -- node[pos=0.5,above,yshift=0.5mm]{$T_{data\;superframe}$}
  (180mm,21mm);

\end{tikzpicture}

%% file: sections/04-TopologyCollection.tex
\section{Topology collection}
\label{sec:topo}

The TDMH topology collection distributed algorithm is one of the key innovations of the protocol, and relies on information from both the clock synchronization and flooding scheme to efficiently gather network topology data and propagate it to the master node.

The topology collection is always running throughout the lifetime of a TDMH network. In this way, the master node is always updated with topology changes and can reschedule streams as needed, as well as being aware of nodes joining/leaving the network. The continuously updated mesh topology is also advantageous for network monitoring and diagnostics, making it possible to preventively add routing nodes in areas of poor connectivity, providing a valuable network troubleshooting tool in an industrial setting.

The current implementation of TDMH is however not targeted at use cases with mobile nodes, as the low bandwidth of the IEEE 802.15.4 PHY layer limits the topology update rate.

\subsection{Network connection}

A node joining a TDMH network needs first to synchronize. To do so, a node needs to receive two clock synchronization frames. The first one allows to establish the offset between the local clock and the network time, while the second one completes the initialization of the FLOPSYNC-2~\cite{bib:TerraneoEtAl-2014a} controller for clock skew compensation. The synchronization allows the joining node to know the boundaries of TDMA slots as well as tiles, and to know its turn in the round-robin access control scheme for control uplink frames. 
The node will then use control uplink frames to participate in the distributed topology collection algorithm to be included in the mesh network, and can also begin opening streams.

It is worth noting that clock synchronization frames are control downlink frames, and thus are flooded using a Glossy-like~\cite{bib:FerrariEtAl-2011a} constructive interference flooding scheme. One important but overlooked characteristic of Glossy is that it provides each node with knowledge of how many hops there are between it and the flood initiator. In TDMH, where floods are always initiated by the master node, we exploit this information to form an efficient convergecast channel forwarding network topology data towards the master node.

Although not necessary for the network operation, the FLOPSYNC implementation of TDMH also allows to perform propagation delay compensation~\cite{bib:TerraneoEtAl-2015a} to endow applications with sub-microsecond clock synchronization, to facilitate real-time distributed applications.

\subsection{Topology connection distributed algorithm}

The topology collection distributed algorithm exploits the broadcast nature of the radio channel. For each control uplink slot one node in turn broadcasts its topology data. The topology data is not flooded, so only a nodes' direct neighbors can receive it. Nodes that overhear this frame can update their local knowledge of the network topology.

This local network connectivity data needs to be forwarded to the master node, in order for it to construct and keep updated the full mesh network graph.

As the master node too overhears control uplink frames, for nodes that are at 1 hop from the master this happens implicitly. Nodes that cannot reach the master directly select a node with a lower hop number as fowardee of their topology data. The forwardee node, upon overhearing the frame will store the forwarded data and later, when its turn to transmit comes, will forward it together with its own data.
This solution exploits the hop information made available by Glossy to guarantee that at every retransmission the topology data is always forwarded closer to the master node, effectively creating a convergecast channel routing topology (and SME) data to the master node in the minimum number of transmissions.

The initial topology collection implementation~\cite{bib:TerraneoEtAl-2018a} only collected a single network graph. However, after an extended experimental evaluation it became evident that certain links could be too weak to be reliably used for data exchange, but at the same time strong enough to cause interference among independent transmissions in the same data slots. For this reason, the topology collection algorithm was extended to collect two network graphs: the graph of strong links, which includes only links that are above a certain RSSI threshold, which is used by the router for stream routing, and the graph of weak links, a superset of the strong links graph, which includes all links regardless of RSSI, which is used by the scheduler as part of the channel spatial reuse for conflict detection.

The information that each node transmits in control uplink slots is the following: its node unique ID, its hop, the node ID of the forwardee, and two bitmasks with its current knowledge of its direct strong and weak neighbors.
Bitmasks are fixed size, requiring a number of bits equal to the maximum number of nodes in the network (a configurable parameter), thus not imposing any limit on the number of neighbors a node can have and preserving the full mesh topology data.
Moreover, a node also transmits forwarded topologies composed of node IDs and bitmasks, as well as SMEs.
The number of forwarded topologies is limited by the control uplink slot size.
In dense networks, a node that is selected as forwardee by more nodes than the available forwarding capability puts forwarded topologies in a queue and sends them in fifo order.

Nodes that are no longer overheard for a configurable number of rounds of the algorithm are removed from the topology, in order to respond to link and node failures.

%% file: sections/05-Scheduler.tex
\section{Router and scheduler}
\label{sec:sched}

The centralized router and scheduler of TDMH is the component that manages all data streams in the network, selecting communication paths and assigning transmissions to individual data slots in order to meet the real-time requirement of applications.
The centralized approach has been selected in TDMH to provide collision-free management of the network without excessive control overhead, as stream opening and closing is assumed a more infrequent operation than data transmission, and can be consequently assigned less bandwidth.

The router and scheduler is run by one node in the network which is assigned the master role. It is expected that the master role is assigned to the network gateway, which connects the network to the Internet or, in an industrial setting, to a Programmable Logic Controller (PLC) or Supervisory Control And Data Acquisition (SCADA) system. In these networks the gateway is already a single point of failure, so a centralized protocol does not add another failure mode.

It should be noted that the router and scheduler are operated on demand. A new schedule is computed and distributed to the network only if new streams are opened or closed, or if the network topology changes in a way that impacts existing streams.

Once a schedule is distributed, given the periodic nature of streams, it is executed in a loop until superseded by a newer schedule.

The router and scheduler is considered a customization point of TDMH, as also done by other protocols such as TSCH.
The current TDMH implementation comes with a fully functional shortest path router and greedy scheduler, but further research is encouraged, such as for load balancing to maximize the network lifetime, which is outside the scope of this paper.

\subsection{Activation logic}

The activation logic is the component in TDMH that decides when a new schedule needs to be computed.

Every time a new schedule is computed, the scheduler produces, other than the schedule itself, two additional data structures used by the activation logic. The first data structure is the set of used links, which is a subset of the graph of strong links, composed of links that are actually used to route streams in the current schedule. The second data structure is the set of spatial reuse conflict links, which is composed entirely of links \emph{not} present in the graph of weak (or strong) links. This is the list of links that, if present, would invalidate the current schedule due to interference in the spatial resue of channels.
These two data strutures are implemented as bloom filters, a memory and computationally efficient set implementation.

Every time the master node receives an uplink frame, the graphs of strong and weak links are updated with both the direct and forwarded topologies. Each link being removed from the graph of strong links is checked against the set of used links. If present, a link currently used to route streams has become unavailable, thus requiring a reschedule. Each link being added to the graph of weak links is instead checked against the set of spatial reuse conflict links. If present, this new link may cause interference among concurrent transmissions, and a reschedule is performed.
Since bloom filters admit false positives, spurious reschedules are possible, if infrequent. This trade-off was considered acceptable compared to the performace advantages of a bloom filter.

In addition, the reception of SME requesting the opening or closing of streams also cause a reschedule.

\subsection{Rounting and scheduling algorithm}

The scheduling algorithm begins by computing the least common multiple against the streams periods in order to compute the schedule duration. It should be noted that the stream API restricts the possible periods of a stream to a logarithmic progression (1,2,5,10,20,50,...) multiples of the tile duration, thus the schedule duration is at most twice the period of the longest stream.

The scheduler implements a greedy incremental algorithm that iterates over all the streams starting from the previously scheduled ones, and for each of them invokes the router to break a stream in individual transmissions over available links. The TDMH router uses a breadth-first search to find the shortest path between the stream endpoints. If the stream has requested the spatial redundancy feature, a limited depth-first search algorithm with depth limit equal to the primary path lenght plus a configurable parameter is used to find a secondary path in the graph.
The scheduler then tentatively allocates each transmission, including those for redundancy, to the first available timeslot, checking for conflicts and advancing to the next time slot as needed. A stream in which all transmissions have been successfully scheduled is accepted and committed to the current schedule, otherwise it is rejected and an IE is sent to notify the endpoints.

The current schedule assumes that the deadline of each stream equals its period, a common assumption in real-time systems~\cite{Hartstone-1992a}. Thus, a stream is accepted if all its transmissions fit within its period, considering conflicts caused by streams scheduled before it.

In the general case of streams with different periods, the streams with shorter period have their transmissions repeated in the schedule, spaced apart exactly by the stream period in order to meet the periodic real-time constraint. This feature makes it possible to perform an efficient conflict check, where only the transmission period and offset from the beginning of the schedule are sufficient to check for conflicts, without the need to check all repetitions. The compact (offset,period) form of the each stream is also the one used for the schedule distribution, greatly reducing the bandwidth requred for the schedule dissemination.

\subsection{Schedule distribution}

Each computed schedule is disseminated through the control downlink flooding. The schedule is transmitted in its compact form, as a set of (source,destination,offset,period) tuples for each transmission, each marked by the stream it belongs to. Each node receiving the schedule only expands the transmissions with a matching source or destination, resulting in an expanded schedule with instructions on what to perform in all schedule slots. During schedule expansion, the allocation of buffers for data frames forwarding takes place.

To provide reliability in case of missed schedule frame reception, each schedule is sent multiple times (3 in the current implementation) before it takes effect. During this time, the previous schedule is executed. In the rare occurrence that a node only has received a partial schedule despite the multiple transmissions, it can request the master to resend it by means of a dedicated SME.

%% file: sections/07-SimScalability.tex
\section{Simulation exploration}
\label{sec:expres}

In this section the TDMH performance is characterized using simulations, while the next section is dedicated to an experimental evaluation using real sensor nodes.
Unless otherwise stated, both in this section and in Section~\ref{sec:experiments}, TDMH was configured with 100ms tiles and 6ms slots, 22\% of which were reserved for control frames.
The networks used for the simulations were all hexagonal-like as in Figure~\ref{fig:comparison-omnet}, where each node has up to six neighbors.

\subsection{Network formation time}

The convergence time of the topology collection algorithm determines the reaction time of TDMH to a topology change, such as a node joining/leaving the network, or links becoming available/unavailable due to environmental changes or interference. A special case is the initial network formation problem, where the entire network topology has to be identified starting from scratch.

Computing the network formation time is difficult to perform in closed form due to the need to take into account the queues of forwarded topologies in every node.
The network formation time mainly depends on the maximum number of nodes for which the network is configured, which defines the round-robin cycle and forwarded topology bitmask sizes, the actual number of nodes in the network, the network topology and the percentage of TDMA slots reserved for control uplink frames.
The OMNeT++ implementation of TDMH can be used to simulate a given network condition and easily compute the network formation time.

\begin{figure}[tb]
 \begin{center}
  \includegraphics[width=0.95\columnwidth]{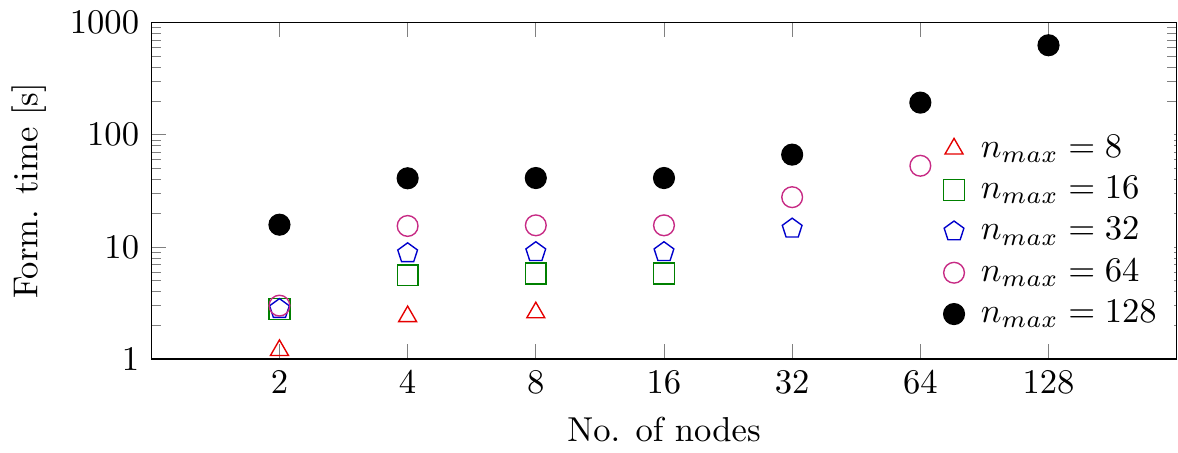}
 \end{center}
 \caption{Network formation time as a function of maximum number of nodes (network configuration parameter) and actual number of nodes.}
 \label{fig:figure_formation_time_sim}
\end{figure}

Figure~\ref{fig:figure_formation_time_sim} shows the network formation time, counted starting when the node clocks are synchronized, and ending when the master has the full graph of the network.
From the figure it can be noted that for networks under 32 nodes as well as for a fully populated 64 node network, the formation time is under 100 seconds. For larger networks it grows up to 629s for a network with 128 nodes. This large convergence time is due to the increase in the bitmask size limiting the number of topologies that can be forwarded in each uplink slot.
To overcome this issue, it is possible to bring the number of uplink frames per uplink tile from 1 to 4. This configuration change reduces the network formation time of a 128 node network to 117s, while increasing the percentage of TDMA slots reserved for network control to only 34\%.

From the simulations presented, it can be concluded that TDMH can easily scale to networks with more than 100 nodes and 10 hops.
Further scaling, e.g. to thousands of nodes, would require a higher data rate PHY layer than the current 250Kbit/s.

\subsection{Power consumption}

The power consumption of TDMH is very predictable. The power consumption of each node due to data transmission can be computed given the current schedule, while that for the control uplink can be computed from the current topology. Finally, the power consumption for control downlink is that required for the flooding of clock syncronization frames, plus the one for the occasional schedule distribution.
\begin{figure}[t]
 \begin{center}
  \includegraphics[width=0.95\columnwidth]{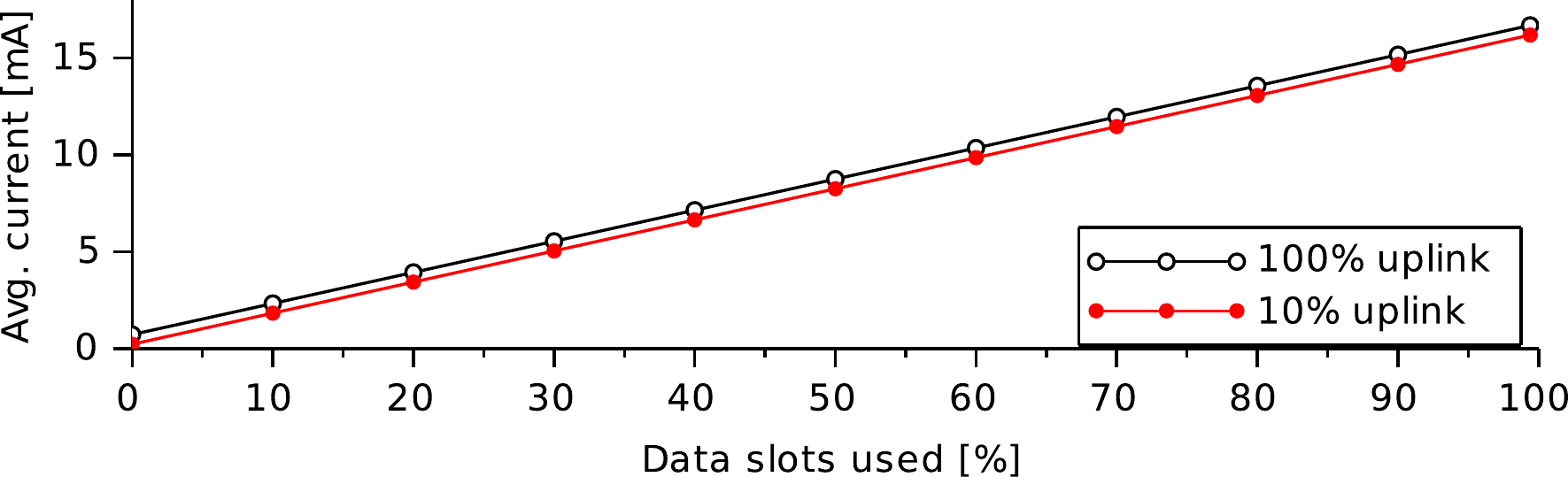}
 \end{center}
 \caption{Average node current consumption as a function of the percentage of data slots used by the schedule, and network connectivity.}
 \label{fig:avgcurrent}
\end{figure}
A simple TDMH power model has been made in which current consumption data for sending, receiving and flooding frames have been measured on WandStem~\cite{bib:Terraneo:2016:DHE:2893711.2893753} nodes, and combined with schedule and topology information from OMNeT++ simulations.

Figure~\ref{fig:avgcurrent} shows the average current consumption of a node running TDMH as a function of the percentage of data slots used for data transmission, and average percentage of uplink slots where a frame is overheard (network connectivity).
The plot clearly shows the power efficiency of TDMH, as the current consumption scales linearly with the data traffic traversing it, as anticipated in Section~\ref{sec:phylayer}. Moreover, denser networks result in only a marginal increase in current consumption due to overhearing uplink frames.

\subsection{Comparison with the state of the art}

\begin{figure}[tb]
 \begin{center}
  \includegraphics[width=0.6\columnwidth]{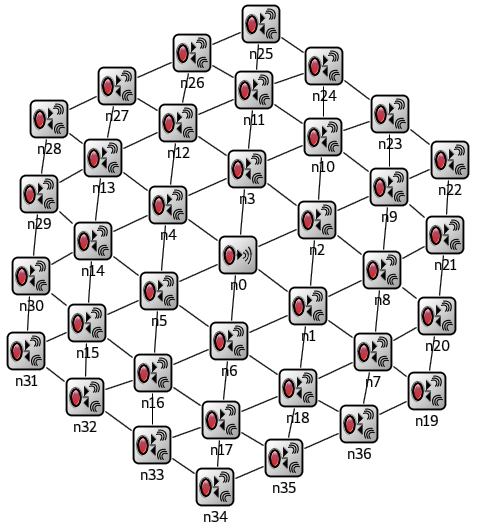}
 \end{center}
 \caption{37 node, 6 hop mesh network simulated in OMNeT++ to compare TDMH with the work by Mager et al~\cite{bib:MagerEtAl-2019a}.}
 \label{fig:comparison-omnet}
\end{figure}

\begin{figure}[t]
 \begin{center}
  \includegraphics[width=0.95\columnwidth]{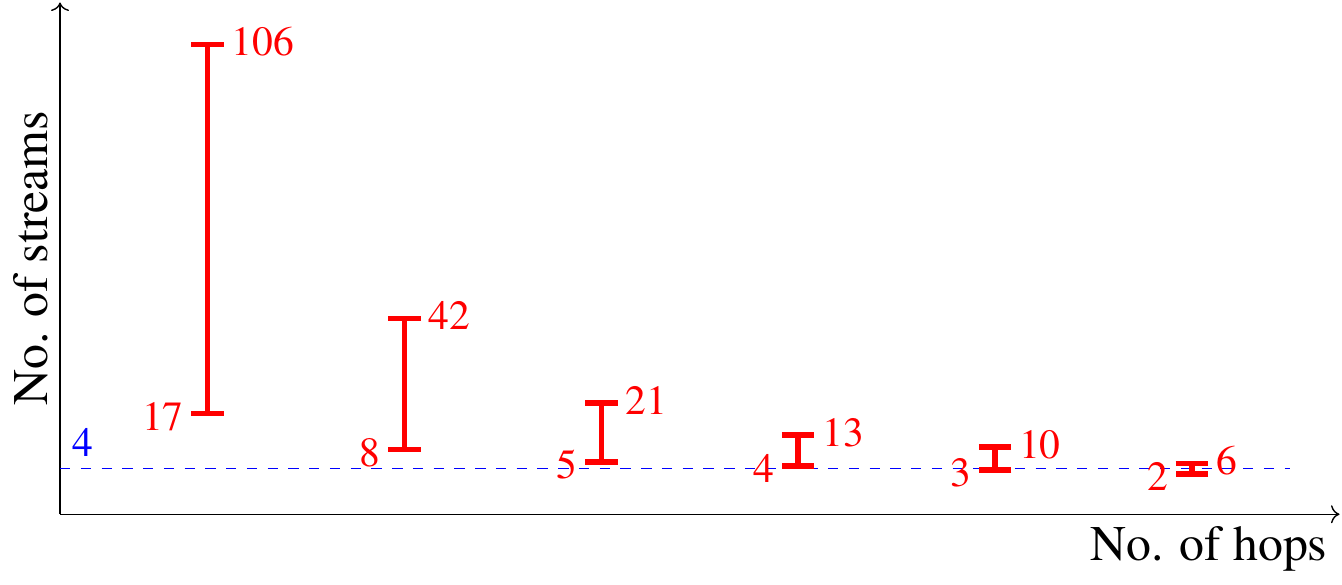}
 \end{center}
 \caption{Number of schedulable streams as a function of the number of hops between endpoints, for TDMH (red ranges) and WCPS (blue dashed line).
 }
 \label{fig:comparison-results}
\end{figure}

Comparing TDMH with other wireless protocols such as TSCH, DSME or LLDN is made difficult by the significant design differences that distinguish it from other protocols, and the corresponding difficulty of establishing common metrics to ground the comparison.

We thus compare TDMH with the Wireless CPS (WCPS) work by Mager et al~\cite{bib:MagerEtAl-2019a}, which shares TDMH's design goal of providing bounded latency multi-hop communication.

As the source code of WCPS is not publicly available, we reimplemented the wireless communication part of the WCPS in OMNeT++, using the Glossy flooding scheme to broadcast data packets following an hardcoded schedule.
We chose to perform the comparison in simulation because the main goal is to compare the two approaches rather than the optimization of the implementations.
As a testbed, we used the network shown in Figure~\ref{fig:comparison-omnet}.

To provide a fair comparison, we configure the OMNeT++ simulation of both TDMH and WCPS with 2ms slots in which a 21 byte message is transmitted across one hop.
This time has been measured experimentally on WandStem nodes for TDMH, and it has been assumed that the original WCPS solution can operate under the same conditions, as no published data is available about the data slot sizing in the WCPS paper.
TDMH streams were opened without redundancy because the WCPS lacks support for redundant transmissions.

The comparison consists in a Monte Carlo simulation selecting random source and destination nodes in the network with a given distance in hops between endpoints and progressively adding them to the current schedule until the specified delay bound of 50ms can no longer be fulfilled. The process is then repeated 200 times with random streams sources/destinations to estimate the minimum and maximum number of schedulable streams in a given configuration.

Figure~\ref{fig:comparison-results} shows how many unidirectional streams can be scheduled by TDMH vs. WCPS, as a function of the number of hops separating the stream endpoints.

The first fact to note is that in the WCPS case, the number of schedulable streams is constant for every distance in hops, and equal to 4.
This is to be expected, as the WCPS approach to multi-hop communication is very different from TDMH, and consists in keeping no topology information and flooding every data packet across the entire network. Thus, WCPS transforms a multi-hop network in a single broadcast and collision domain. The lack of topology information means that, regardless of the physical location of the stream endpoints in the network, packets need to be flooded for a number of hops equal to maximum number of hops in the network, thus occupying the same number of data slots.

On the other hand, the number of schedulable streams by TDMH increases as the number of hops between endpoints decreases, as TDMH can schedule transmissions using the minimum number of slots required to connect the endpoints. Moreover, due to the spatial reuse of channels, TDMH can even schedule more than one transmission in the same time slot if said transmissions do not interfere. For these reasons, TDMH can schedule up to 106 streams.
It should be noted that for TDMH the number of schedulable streams is not a single number, but a range. This is because during the Monte Carlo simulation, a random selection of stream sources and destinations causes different conflict patterns.
Another fact to note is that for large hop distances between endpoints, TDMH can sometines schedule less streams than WCPS.
This fact is due to TDMH needing to reserve some TDMA slots for for network control.
This overhead is however what makes it possible for TDMH to open new streams at runtime, while in the current WCPS solution as published in~\cite{bib:MagerEtAl-2019a} any change in the streams would require to manually update the firmware of all the nodes. Adding schedule distribution at runtime to WCPS would thus introduce a similar overhead.

A few more notes conclude this comparison.
First, in a real-world implementation of WCPS the actual number of hops of the network may be hard to estimate and may also be time varying due to links becoming unavailable, --a matter not discussed in the original paper--. Although we did not take this phenomenon into account in the simulation, in a real-world WCPS deployment the maximum hop numer a packet needs to be flooded needs to be oversized to prevent temporary loss of connectivity due to link failures causing the rebroadcast count to become insufficient, further reduing the number of schedulable streams compared to our simulation.
TDMH employs flooding too, but only for clock synchronization and schedule distribution, thus oversizing does not cause a penalty for every data transmissions.

Finally, we believe that both solutions have their place. The WCPS approach can support moving nodes, while --at least with the bandwdith limitations of the current PHY layer-- TDMH cannot react fast enough to topology changes to support mobile nodes. On the other hand, TDMH supports opening and closing streams at runtime, and can provide a significant bandwidth increase thanks to its spatial channel reuse and mesh network topology knowlege, as well as providing network monitoring and diagnostic information unavailable in WCPS.

%% file: sections/08-ExpRes.tex
\section{Experimental evaluation}
\label{sec:experiments}

TDMH has been validated experimentally on real WSN nodes, where constructive interference has to occur also in the presence of multipaths, transmissions are affected by interference, and clock synchronization has to track the time varying drift of quartz crystals.
For the experiments we used a testbed composed of 14 WandStem WSN nodes distributed in the first floor of building 21 of the Politecnico di Milano. The placement of the nodes is shown in Figure~\ref{fig:ExperimentTopology} and is the same for all the reported experiments. The same figure also shows the topology as collected by the topology collection during one of the experiments. Slight variations in the topology were observed between experiments, and also within experiments, the latter case causing reschedules if the changes affects open streams. This is to be expected, as interference, as well as people and objects moving in the environment introduce link strength variations.

Without loss of generality, the experiments were performed by opening a stream from each node towards the master node (node 0 in the figure), as the master was the only node connected to a computer for logging purposes.
This experimental setup allowed to collect network statistics including the reliability of each stream by monitoring only the master node logs.

\subsection{Real-time communication and reliability}

\begin{table*}[tb]
{
    \vspace{0.4em}
    \caption{Stream reliability experimental evaluation.}
    \vspace{0.4em}
    \small
     \begin{center}
      \begin{tabular}{|l|r|r|r|l|r|r|r|}
      \hline
      Experiment & \multicolumn{3}{c|}{Network}       & \multicolumn{2}{c|}{Best stream(s)}  & \multicolumn{2}{c|}{Worst stream(s)} \\
      \hline
                 & \# Packets & Avg. reliability & \# Schedules & SRC(s)           & Reliability       & SRC  & Reliability \\
      \hline
      Standard 1 & 776780 & 99.99\% & 19 & 1,5,6,13         & 100.00\% &  9 & 99.94\% \\
      Standard 2 & 773756 & 99.98\% & 23 & 1,5,6,7,8,12     & 100.00\% & 10 & 99.83\% \\
      Standard 3 & 774816 & 99.98\% & 30 & 1,2,3,4,6,7,8,9  & 100.00\% & 13 & 99.88\% \\
      \hline
      Interf.  1 & 261031 & 99.58\% & 55 & 6                &  99.98\% & 11 & 99.15\% \\
      Interf.  2 & 304593 & 99.55\% & 51 & 6                &  99.99\% &  4 & 99.11\% \\
      Interf.  3 & 258629 & 99.82\% & 62 & 1,6              & 100.00\% &  5 & 99.14\% \\
      \hline
      \end{tabular}
      \label{tab:streamreliability}
     \end{center}
}
\end{table*}

To collect information about the reliability of TDMH streams six experiments were made, for a total of more than 3 million data packets being exchanged between stream endpoints. As the streams were opened with triple redundancy, and as most streams require multiple hops to connect endpoints, the actual number of data frames being transmitted exceeds 10 million.

The six experiments were divided in two categories, the standard tests and tests with interference. In the standard tests no additional interference sources other than the building's Wi-Fi infrastructure were present, while in the interference case throughout the entire experiment  web surfing traffic was generated from a Wi-Fi access point placed close to node 0 and configured for a 20dBm output power (while TDMH nodes were transmitting at 5dBm output power). In addition, during the interference tests, two microwave ovens were shortly turned on close to nodes 3 and 4. The microwave ovens were shown to affect weak links in nearby nodes and consequently adding to the interference level as well as introduce topology changes and reschedules.

During all six experiments, the scheduler was always able to schedule all streams in the available slots with a latency lower than the stream period, without the need to close streams due to network capacity exhaustion. The real-time communication requirements were thus met for all streams.

Experimental results are summarized in Table~\ref{tab:streamreliability}. The first part of the table shows the aggregated total number of packets being transmitted by all streams and the aggregated reliability for all streams. Reliability was computed as the percentage of transmitted packets that were correctly received at the other endpoint. The total number of schedules being computed during the experiment is also reported. This number includes the schedules required for the initial network formation.

The second part of the table shows the statistics for the best and worst stream of each experiment. 
As each node opens one stream towards the master node, the SRC number uniquely identifies the stream. For example, 1 means the stream 1 $\rightarrow$ 0, opened between node 1 and node 0.
In most experiments more than one stream achieved the same reliability, and in such a case the list of all streams with the same reliability level is reported.

As can be seen, TDMH can achieve greater than 99.9\% reliability while providing bounded-latency real-time communication. The presence of external interferences in the environment where TDMH is operated causes packet losses and link unreliability that triggers additional communications rescheduling, but the reliabilty of data streams is maintenied above 99.5\%.

\subsection{Effect of redundancy}

\begin{table}[htb]
{
    \vspace{0.4em}
    \caption{Effect of redundancy on stream reliability.}
    \vspace{0.4em}
    \small
     \begin{center}
      \begin{tabular}{|l|r|r|r|}
      \hline
      Experiment & No redundancy & Double red. & Triple red. \\
      \hline
      Standard   & 99.71\% & 99.95\% & 99.98\% \\
      \hline
      Interf.    & 97.11\% & 99.10\% & 99.65\% \\
      \hline
      \end{tabular}
      \label{tab:streamredundancy}
     \end{center}
}
\end{table}

\begin{figure}[t]
 \begin{center}
  \includegraphics[width=0.95\columnwidth]{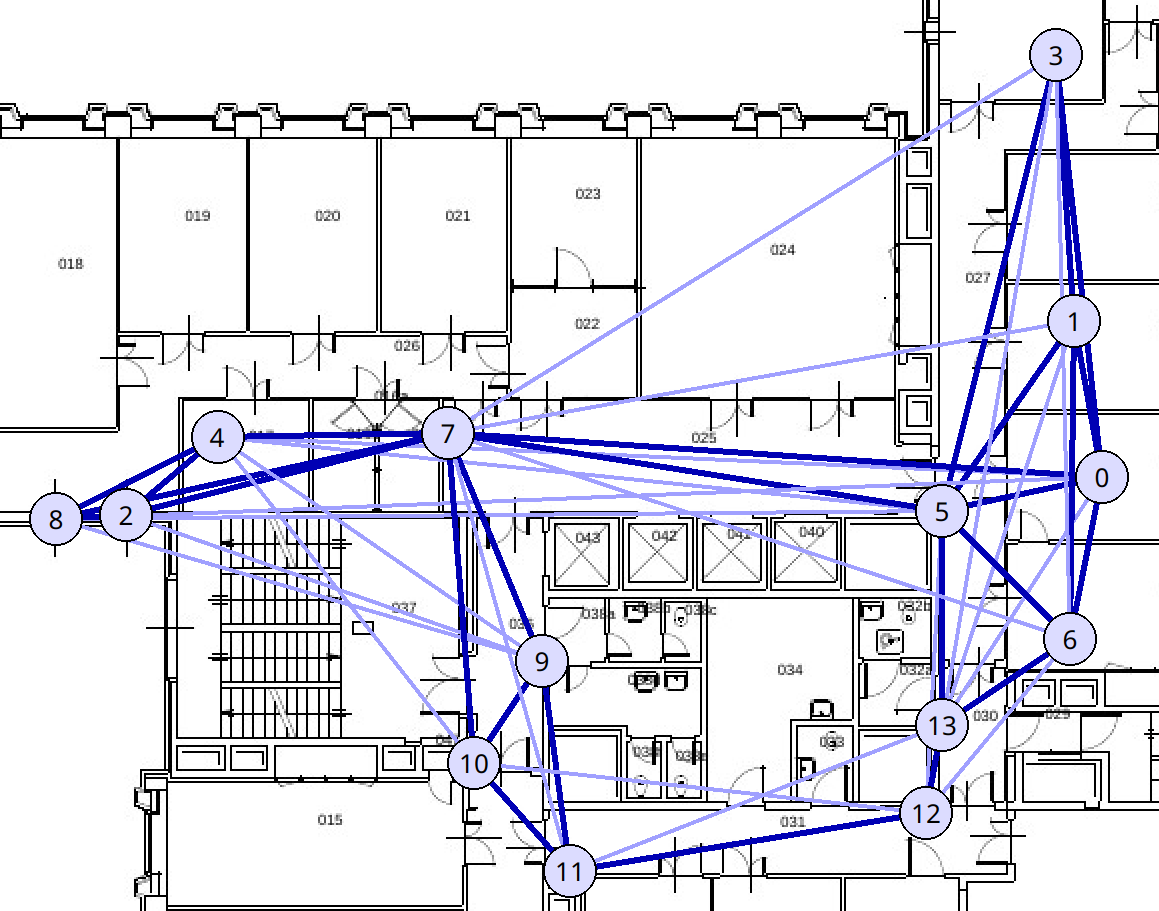}
 \end{center}
 \caption{Node placement for the experimental evaluation, showing the node ID and the network graph produced by the topology collection (strong links in dark blue, weak links in light blue).}
 \label{fig:ExperimentTopology}
\end{figure}

The effect of redundancy on stream reliability was also studied.
Table~\ref{tab:streamredundancy} shows the aggregated reliability for the standard and interference experiments considering only one transmission per period (no redundancy), two (double redundancy) or three transmissions (triple redundancy).

As can be seen, the average reliability of streams without redundancy is above 99.7\% without interference, and 97\% with interference. The highest reliability gain occurs between the no redundancy and single redundancy case, while moving to triple redundancy affords a comparatively lower reliability increase.
Redundancy was experimentally shown to provide a beneficial effect to the reliability of streams, bringing it, on average, above 99.6\% even in the presence of interferences.

\subsection{Effect of node failures}

An experiment is here reported to demonstrate the behavior of TDMH streams in the presence of a node failure, showing how the combined effect of spatial redundancy and mesh network topology monitoring can help provide reliable communication.

\begin{figure}[t]
  \includegraphics[width=0.7125\columnwidth]{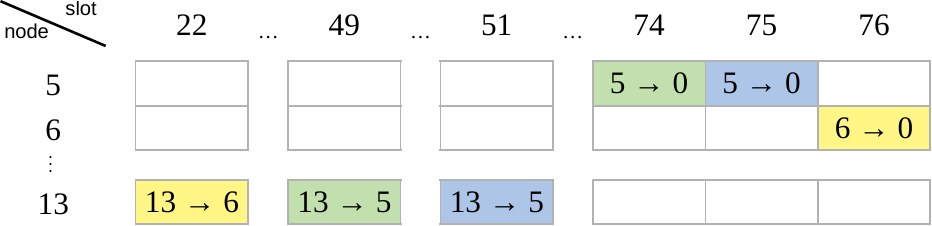}
  
  \vspace{5mm}
  
  \includegraphics[width=1.00\columnwidth]{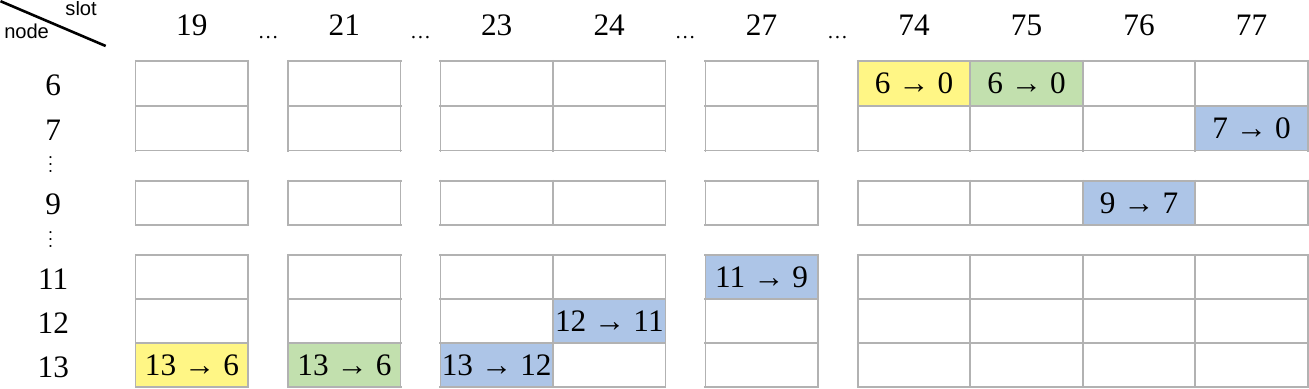}
 \caption{Schedule fragment showing how the stream 13 $\rightarrow$ 0 is scheduled before (top) and after (bottom) the failure of node 5.}
 \label{fig:nodeFailureSchedules}
\end{figure}

The stream under test is 13 $\rightarrow$ 0. Before the fault introduction, the network topology is as shown in Figure~\ref{fig:ExperimentTopology}, and the corresponding schedule is the one shown in the top part of Figure~\ref{fig:nodeFailureSchedules}.

The stream 13 $\rightarrow$ 0 is configured for triple redundancy with the spatial redundancy feature activated. Thus, whenever possible, the transmissions between the endpoints follow two paths in the mesh network. As can be seen from the schedule, the three redundant transmissions follow the path 13 $\rightarrow$ 6 $\rightarrow$ 0 once, and 13 $\rightarrow$ 5 $\rightarrow$ 0 twice. The ordering of the transmissions in the schedule and the gaps in the slot numbers exist because the scheduler has to schedule 12 other streams and has to consider transmission conflicts.

At a certain time instant during the experiment that we will call T+0.0 seconds, node 5 was turned off by removing its batteries. As a result, two of the three redundant transmissions of node 13 start failing systematically, but due to the spatial redundancy, the packets are still received through the path traversing node 6.

Neighboring nodes start no longer overhearing node 5 during the control uplink slots and after the prescribed timeout remove it from their topologies. At T+28.1 seconds the master node notices the topology change and begins rescheduling.

The new schedule is shown at the bottom of Figure~\ref{fig:nodeFailureSchedules}. As can be seen, despite the failure of node 5, the scheduler could still find two paths in the mesh network: 13 $\rightarrow$ 6 $\rightarrow$ 0 and 13 $\rightarrow$ 12 $\rightarrow$ 11 $\rightarrow$ 9 $\rightarrow$ 7 $\rightarrow$ 0. Thus, two of the three redundant transmissions are scheduled through the shortest path, and the third one through the longer one preserving the spatial redundancy.

At T+36.1 seconds, after the schedule has been computed and disseminated, all nodes in the network switch to the new schedule, and the redundancy in the packet transmission of the stream 13 $\rightarrow$ 0 is reestablished. During this time frame, thanks to the spatial redundancy, stream 13 $\rightarrow$ 0 lost no packets.

In the general case, temporary loss of redundant transmissions due to node failures increase the susceptibility to external interference until redundancy is reestablished through rescheduling.

%% file: sections/09-Conclusions.tex
\section{Conclusions and future work}
\label{sec:conclusions}

This paper presented TDMH, a wireless mesh network stack designed from the ground up for real-time applications.
It is expected that this protocol will be used for real-time distributed applications, and the scheduler customization point will foster additional research from the real-time community.

%% file: 2020-TimeDetWirelessStack-Preprint.bbl
\begin{thebibliography}{100}

\bibitem{seiler2001analysis}
Pete Seiler and Raja Sengupta.
\newblock Analysis of communication losses in vehicle control problems.
\newblock In {\em Proc. 2001 American Control Conference}, volume~2, pages
  1491--1496, Arlington, VA, USA, 2001. IEEE.

\bibitem{4560038}
B.~{Cheng}, M.~{Yuksel}, and S.~{Kalyanaraman}.
\newblock {Orthogonal Rendezvous Routing Protocol for Wireless Mesh Networks}.
\newblock {\em IEEE/ACM Transactions on Networking}, 17(2):542--555, 2009.

\bibitem{802154e-survey}
Domenico De~Guglielmo, Simone Brienza, and Giuseppe Anastasi.
\newblock {IEEE 802.15.4e: A survey}.
\newblock {\em Computer Communications}, 88:1--24, 2016.

\bibitem{5418856}
J.~{Luo}, C.~{Rosenberg}, and A.~{Girard}.
\newblock {Engineering Wireless Mesh Networks: Joint Scheduling, Routing, Power
  Control, and Rate Adaptation}.
\newblock {\em IEEE/ACM Transactions on Networking}, 18(5):1387--1400, 2010.

\bibitem{wirelesshart}
Wirelesshart™, 2018-02.

\bibitem{treemac}
W.~Z. Song, R.~Huang, B.~Shirazi, and R.~LaHusen.
\newblock {TreeMAC: Localized TDMA MAC protocol for real-time high-data-rate
  sensor networks}.
\newblock In {\em 2009 IEEE International Conference on Pervasive Computing and
  Communications}, pages 1--10, 2009.

\bibitem{packmac}
K.~Moriyama and Y.~Zhang.
\newblock {An Efficient Distributed TDMA MAC Protocol for Large-Scale and
  High-Data-Rate Wireless Sensor Networks}.
\newblock In {\em 2015 IEEE 29th International Conference on Advanced
  Information Networking and Applications}, pages 84--91, 2015.

\bibitem{bib:Terraneo:2016:DHE:2893711.2893753}
Federico Terraneo, Alberto Leva, and William Fornaciari.
\newblock Demo: A high-performance, energy-efficient node for a wide range of
  wsn applications.
\newblock In {\em Proceedings of the 2016 International Conference on Embedded
  Wireless Systems and Networks}, EWSN '16, pages 241--242, USA, 2016. Junction
  Publishing.

\bibitem{7005204}
G.~Patti, G.~Alderisi, and L.~L. Bello.
\newblock {Introducing multi-level communication in the IEEE 802.15.4e
  protocol: The MultiChannel-LLDN}.
\newblock In {\em Proceedings of the 2014 IEEE Emerging Technology and Factory
  Automation (ETFA)}, pages 1--8, 2014.

\bibitem{Hartstone-1992a}
N.H. Weiderman and N.I. Kamenoff.
\newblock Hartstone uniprocessor benchmark: definitions and experiments for
  real-time systems.
\newblock {\em Real-Time Syst.}, 4(4):353--382, 1992.

\bibitem{4453854}
V.~C. {Gungor}, Ö.~B. {Akan}, and I.~F. {Akyildiz}.
\newblock {A Real-Time and Reliable Transport (RT)$^{2}$ Protocol for Wireless
  Sensor and Actor Networks}.
\newblock {\em IEEE/ACM Transactions on Networking}, 16(2):359--370, 2008.

\bibitem{muchmac}
Joris Borms, Kris Steenhaut, and Bart Lemmens.
\newblock Low-overhead dynamic multi-channel mac for wireless sensor networks.
\newblock In Jorge~S{\'a} Silva, Bhaskar Krishnamachari, and Fernando Boavida,
  editors, {\em Wireless Sensor Networks}, pages 81--96, Berlin, Heidelberg,
  2010. Springer Berlin Heidelberg.

\bibitem{bib:TerraneoEtAl-2014a}
F.~Terraneo, L.~Rinaldi, M.~Maggio, A.~V. Papadopoulos, and A.~Leva.
\newblock {FLOPSYNC-2}: Efficient monotonic clock synchronisation.
\newblock RTSS, pages 11--20, 2014.

\bibitem{bib:TerraneoEtAl-2018a}
F.~Terraneo, P.~Polidori, A.~Leva, and W.~Fornaciari.
\newblock {TDMH-MAC}: Real-time and multi-hop in the same wireless {MAC}.
\newblock In {\em Proc. 39th IEEE Real-Time Systems Symposium}, Nashville, TN,
  USA, 2018.

\bibitem{7389381}
G.~Tian, S.~Camtepe, and Y.~Tian.
\newblock A deadline-constrained 802.11 mac protocol with qos differentiation
  for soft real-time control.
\newblock {\em IEEE Transactions on Industrial Informatics}, 12(2):544--554,
  April 2016.

\bibitem{5678602}
K.~{Kim} and K.~G. {Shin}.
\newblock {Self-Reconfigurable Wireless Mesh Networks}.
\newblock {\em IEEE/ACM Transactions on Networking}, 19(2):393--404, 2011.

\bibitem{WirelessHart-closedloop}
A.~{Santos}, D.~{Lopes}, J.~{César}, L.~{Luciano}, A.~{Neto}, L.~A. {Guedes},
  and I.~{Silva}.
\newblock Assessment of wirelesshart networks in closed-loop control system.
\newblock In {\em 2015 IEEE International Conference on Industrial Technology
  (ICIT)}, pages 2172--2177, 2015.

\bibitem{1709950}
Y.~{Huang}, R.~{Guerin}, and P.~{Gupta}.
\newblock {Supporting Excess Real-Time Traffic With Active Drop Queue}.
\newblock {\em IEEE/ACM Transactions on Networking}, 14(5):965--977, 2006.

\bibitem{bib:MagerEtAl-2019a}
Fabian Mager, Dominik Baumann, Romain Jacob, Lothar Thiele, Sebastian Trimpe,
  and Marco Zimmerling.
\newblock {Feedback Control Goes Wireless: Guaranteed Stability over Low-power
  Multi-hop Networks}.
\newblock In {\em Proceedings of the 10th ACM/IEEE International Conference on
  Cyber-Physical Systems}, ICCPS '19, pages 97--108, New York, NY, USA, 2019.
  ACM.

\bibitem{7945906}
V.~Díez, A.~Arriola, I.~Val, and M.~Vélez.
\newblock Validation of rf communication systems for industry 4.0 through
  channel modeling and emulation.
\newblock In {\em 2017 IEEE International Workshop of Electronics, Control,
  Measurement, Signals and their Application to Mechatronics (ECMSM)}, pages
  1--6, May 2017.

\bibitem{whartvsisa}
S.~Petersen and S.~Carlsen.
\newblock {WirelessHART Versus ISA100.11a: The Format War Hits the Factory
  Floor}.
\newblock {\em IEEE Industrial Electronics Magazine}, 5(4):23--34, 2011.

\bibitem{ISA-100.11a}
Isa-100.11a, 2018-02.

\bibitem{7993826}
T.~Karimireddy and S.~Zhang.
\newblock Guaranteed timely delivery of control packets for reliable industrial
  wireless networks in industry 4.0 era.
\newblock In {\em 2017 Ninth International Conference on Ubiquitous and Future
  Networks (ICUFN)}, pages 456--461, July 2017.

\bibitem{wsnmacsurvey}
P.~Huang, L.~Xiao, S.~Soltani, M.~W. Mutka, and N.~Xi.
\newblock {The Evolution of MAC Protocols in Wireless Sensor Networks: A
  Survey}.
\newblock {\em IEEE Communications Surveys Tutorials}, 15(1):101--120, 2013.

\bibitem{bib:TerraneoEtAl-2015a}
Federico Terraneo, Alberto Leva, Silvano Seva, Martina Maggio, and
  Alessandro~Vittorio Papadopoulos.
\newblock Reverse flooding: Exploiting radio interference for efficient
  propagation delay compensation in {WSN} clock synchronization.
\newblock In {\em 2015 {IEEE} Real-Time Systems Symposium, {RTSS} 2015, San
  Antonio, Texas, USA, December 1-4, 2015}, pages 175--184, 2015.

\bibitem{mmsn}
G.~Zhou, C.~Huang, T.~Yan, T.~He, J.~A. Stankovic, and T.~F. Abdelzaher.
\newblock Mmsn: Multi-frequency media access control for wireless sensor
  networks.
\newblock In {\em Proceedings IEEE INFOCOM 2006. 25TH IEEE International
  Conference on Computer Communications}, pages 1--13, April 2006.

\bibitem{6841643}
L.~{Yang}, Y.~E. {Sagduyu}, J.~{Zhang}, and J.~H. {Li}.
\newblock {Deadline-Aware Scheduling With Adaptive Network Coding for Real-Time
  Traffic}.
\newblock {\em IEEE/ACM Transactions on Networking}, 23(5):1430--1443, 2015.

\bibitem{8336984}
P.~C. Bartolomeu, M.~Alam, J.~C. Ferreira, and J.~Fonseca.
\newblock Supporting deterministic wireless communications in industrial iot.
\newblock {\em IEEE Transactions on Industrial Informatics}, pages 1--1, 2018.

\bibitem{ISA100.11a-closedloop}
Guilherme Bertelli, Anderson Santos, Júlio César, and Ivanovitch Silva.
\newblock {Performance Evaluation of ISA100.11a Wireless Feedback Control}.
\newblock {\em IFAC-PapersOnLine}, 49(30):290 -- 295, 2016.
\newblock 4th IFAC Symposium on Telematics Applications TA 2016.

\bibitem{8355905}
L.~Leonardi, G.~Patti, and L.~L. Bello.
\newblock Multi-hop real-time communications over bluetooth low energy
  industrial wireless mesh networks.
\newblock {\em IEEE Access}, pages 1--1, 2018.

\bibitem{6728869}
Y.~H. Wei, Q.~Leng, S.~Han, A.~K. Mok, W.~Zhang, and M.~Tomizuka.
\newblock Rt-wifi: Real-time high-speed communication protocol for wireless
  cyber-physical control applications.
\newblock In {\em 2013 IEEE 34th Real-Time Systems Symposium}, pages 140--149,
  Dec 2013.

\bibitem{trama}
Venkatesh Rajendran, Katia Obraczka, and J.~J. Garcia-Luna-Aceves.
\newblock {Energy-efficient Collision-free Medium Access Control for Wireless
  Sensor Networks}.
\newblock In {\em Proceedings of the 1st International Conference on Embedded
  Networked Sensor Systems}, SenSys '03, pages 181--192, New York, NY, USA,
  2003. ACM.

\bibitem{OpenDSME}
Maximilian K{\"o}stler, Florian Kauer, Tobias L{\"u}bkert, Volker Turau,
  J~Scholz, and A~von Bodisco.
\newblock {Towards an Open Source Implementation of the IEEE 802.15.4 DSME Link
  Layer}.
\newblock {\em Proceedings of the 15. GI/ITG KuVS Fachgespr{\"a}ch Sensornetze,
  J. Scholz and A. von Bodisco, Eds. University of Applied Sciences Augsburg,
  Dept. of Computer Science}, 2016.

\bibitem{bib:FerrariEtAl-2011a}
F.~Ferrari, M.~Zimmerling, L.~Thiele, and O.~Saukh.
\newblock Efficient network flooding and time synchronization with {Glossy}.
\newblock IPSN, pages 73--84, 2011.

\end{thebibliography}
